\documentclass[aps,prl,twocolumn,showpacs,groupedaddress,superscriptaddress,footinbib,longbibliography]{revtex4-2}

\usepackage{mathrsfs}
\usepackage{natbib,hyperref}
\setcitestyle{square,sort&compress,comma,numbers}
\hypersetup{colorlinks=true, citecolor=blue, urlcolor=blue, linkcolor=blue}
\usepackage[english]{babel}
\usepackage[utf8]{inputenc}
\usepackage{graphicx}
\usepackage[caption=false]{subfig}
\usepackage{amsmath}
\usepackage{amsfonts}
\usepackage{amssymb}
\usepackage{color,soul}
\setulcolor{red}
\usepackage{easyReview}
\usepackage{xcolor}
\usepackage{tikz}
\usepackage{pgfplots}
\usepgfplotslibrary{colormaps}
\pgfplotsset{
    colormap={mycolormap}{
        rgb255=(59,76,192)
        rgb255=(255,255,255)
        rgb255=(180,4,38)
    }
}

\newcommand\diff{\mathrm{d}}
\renewcommand{\vec}[1]{\mathbf{#1}}

\begin{document}

\title{Flow Through Porous Media at the Percolation Transition}

\author{Mirko Residori}
\affiliation{Institute of Scientific Computing, Technische Universit\"at Dresden, 01062 Dresden, Germany}
\affiliation{Max Planck Institute for the Physics of Complex Systems, N\"othnitzer Stra{\ss}e 38,
01187 Dresden, Germany}
\author{Suvendu Mandal}
\affiliation{Technische Universit\"at Darmstadt, Karolinenplatz 5, 64289 Darmstadt, Germany}
\author{Axel Voigt}
\affiliation{Institute of Scientific Computing, Technische Universit\"at Dresden, 01062 Dresden, Germany}
\affiliation{Center for Systems Biology Dresden, Pfotenhauerstraße 108, 01307 Dresden, Germany}
\affiliation{Cluster of Excellence, Physics of Life, TU Dresden, Arnoldstraße 18, 01307 Dresden, Germany}
\author{Christina Kurzthaler}
\email{ckurzthaler@pks.mpg.de}
\affiliation{Max Planck Institute for the Physics of Complex Systems, N\"othnitzer Stra{\ss}e 38,
01187 Dresden, Germany}
\affiliation{Center for Systems Biology Dresden, Pfotenhauerstraße 108, 01307 Dresden, Germany}
\affiliation{Cluster of Excellence, Physics of Life, TU Dresden, Arnoldstraße 18, 01307 Dresden, Germany}

\begin{abstract} 
We study low-Reynolds-number fluid flow through a two-dimensional porous medium modeled as a Lorentz gas. Using extensive finite element simulations we fully resolve the flow fields for packing fractions approaching the percolation threshold. Near the percolation transition, we find a power-law scaling of the flow rate versus the pressure drop with an exponent of $\approx 5/2$, which has been predicted earlier by mapping the macroscopic flow to a discrete flow network [{\it Phys. Rev. Lett.} {\bf 54}, 1985]. Importantly, we observe a rounding of the scaling behavior at small system sizes, which can be rationalized via a finite-size scaling ansatz. Finally, we show that the distribution of the kinetic energy exhibits a power-law scaling over several decades at small energies, originating from collections of self-similar, viscous eddies in the dead-end-channels. Our results lay the foundation for unraveling critical behavior of complex fluids omnipresent in biological and geophysical systems. 
\end{abstract} 

\maketitle

Transport processes through complex media play fundamental roles in diverse areas, ranging from ancient irrigation systems~\cite{rost2022irrigation} to modern technological applications, such as oil recovery from tight reservoirs~\cite{Gao:2011}, water filtration~\cite{Song:2023} and processing of polymeric materials~\cite{Datta:2022}, to geophysics~\cite{Garcia:2008,voigtlaender2023soft}, where pollutants contaminate rocky materials, to biology. The latter involves the circulation of blood through microvascular networks~\cite{viallat:2019},
microbial motion through wet soil and dense tissues~\cite{Wioland:2013, Wioland:2016, Secchi:2016,Doostmohammadi:2018, Bhattacharjee:2019,Alonso:2019,Kurzthaler:2021,Codutti:2022, Keogh:2024, Scheidweiler:2024, Jin:2023}, and molecular transport in the interior of crowded cells~\cite{Hofling:2013} and unicellular organisms~\cite{Woodhouse:2013,Goldstein:2015, Alim:2017:physaris}, to name a few. In these heterogeneous environments, fluid flow encounters highly disordered porous structures, yielding numerous regions of stagnant flow alongside a few pores of rapid fluid streams \cite{Andrea:1997, Scholz:2012,Alim:2017,deAnna:2017, Meigel:2022,Bordoloi:2022,Residori:2023}. These intricate flow fields can generate hydrodynamic dispersion of individual consitutents~\cite{Taylor:1953, Saffman:1960, Brenner:1980, DEJOSSELINDEJONG:1972} and are strongly affected by complex fluid properties~\cite{Datta:2022, Kumar:2022}. Thus, the interplay of hydrodynamic flows and densely-packed, disordered environments represents an exciting topic at the interface of fluid mechanics and statistical physics with implications for various fields, yet a firm understanding of flow through dense porous media is still lacking even in the Stokes regime. 

In the realm of low-Reynolds-number flows, the behavior of viscous fluids traversing porous media is conventionally captured by Darcy's law. This phenomenological law establishes a relationship, valid at length scales vastly exceeding the pore structure, between the average fluid velocity $\langle \vec{u}\rangle$, the pressure gradient $\nabla p$ imposed across the porous medium, and the scalar permeability $k$ alongside fluid viscosity $\mu$: $\mu\langle\vec{u}\rangle= -k \nabla p$. While anisotropic features of the porous medium can be incorporated in a permeability tensor $\vec{K}$~\cite{Evans:2015}, replacing $k$, such a continuum description tends to overlook the intricate details of the local pore structure. Recent experimental investigations, particularly within dense environments, have illuminated the critical role of strong spatial variations in fluid flow~\cite{Scholz:2012, Datta:2013, Alim:2017}, potentially exerting a profound influence on the macroscopic flow rate. 

Important insights into the underlying physics could come from percolation theory~\cite{Torquato:book}, which deals with system properties that emerge as the density of obstacles increases to a critical packing fraction $\phi_c$, where a sample-spanning cluster first appears. At this percolation threshold, observables are predicted to exhibit power-law scaling behaviors, akin to classical phase transitions. Importantly, a scaling theory predicts a power-law behavior of the flow rate of Stokes flow through a porous medium, modeled as a Lorentz gas, near the critical packing fraction: $Q~\sim (\phi_c-\phi)^\alpha$ with $\alpha\approx 5/2$ in two dimensions (2D) and $\alpha\approx4.4$ in three dimensions (3D)~\cite{Halperin:1985}. This scaling prediction relies on mapping the complex flow through the Lorentz gas onto a flow network by accounting for its geometrical features. In contrast to phenomenological laws, this would provide fundamental physical insights from first principles, but has not yet been confirmed, neither numerically nor experimentally.

\begin{figure*}[tp]
\includegraphics[width=\linewidth]{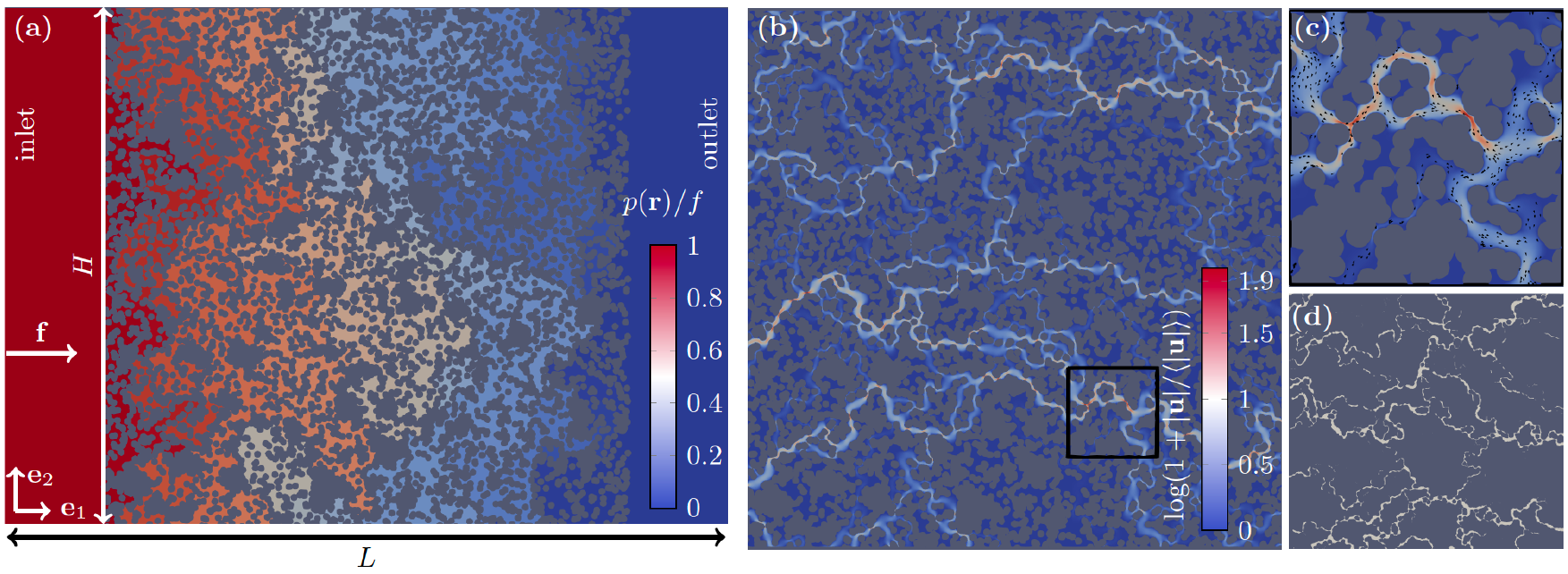}
\caption{(a) Model set-up. A porous medium, modeled by overlapping discs in a domain of size $H\times H$, is placed at the center of a channel of length~$L$ and width~$H$.  Fluid is pushed at a force $\vec{f}$ from the channel inlet to the outlet. The background color indicates the pressure field $p(\vec{r})/f$. Here, the width and length are $H=150a$ and $L = 1.4H$, respectively, where $a$ denotes the disc radius. The packing fraction of overlapping discs is $\phi\approx 0.92\phi_c$.  Disconnected void space is shown in grey. (b-c)~Velocity fields. (b)~Rescaled magnitude of the velocity $|\vec{u}| / \langle|\vec{u}|\rangle$ shown in a logarithmic scale, where $\langle|\vec{u}|\rangle$ is the velocity magnitude averaged over the connected part of the sample. (c)~Zoom of the black frame in (b). Black spikes indicate the flow direction. (d) Backbone of the velocity field, where regions of velocities $\ge \langle|\vec{u}|\rangle$ are indicated in light grey. 
\label{fig:set-up}}
\end{figure*}  

Here, we compute the flow field of a viscous fluid passing through a 2D porous medium, described by the Lorentz gas, for a wide range of packing fractions using extensive finite element simulations. Our results reveal a power-law behavior of the flow rate near the percolation threshold with exponent $\approx 5/2$, thereby, confirming for the first time the prediction by Halperin et al.~\cite{Halperin:1985}. Importantly, these findings demonstrate that for dense porous media the macroscopic flow rate is effectively determined by flow through a discrete random network of the form of the `nodes-links-blob' model describing the geometric structure of the Lorentz gas close to percolation. For small systems the power-law behavior disappears and the data can be collapsed via a finite-size scaling ansatz with theoretically predicted exponents as input. Further, we find a power-law scaling of the distribution of the kinetic energy close to percolation that can be rationalized by collections of viscous, self-similar eddies omnipresent in the dead-end-channels of the stagnant zones.  

\paragraph{Model.--} We consider incompressible, viscous flow through a porous medium in 2D. The quasi-steady, spatially-dependent fluid velocity $\vec{u}(\vec{r})$ and pressure fields~$p(\vec{r})$ are described by the Stokes and continuity equations,
\begin{align}
\mu \nabla^2 \vec{u} - \nabla p + \vec{f}=\vec{0} \quad \text{and} \quad
\nabla\cdot\vec{u} = 0, \label{eq:stokes}
\end{align}
where $\mu$ denotes the fluid viscosity and $\vec{f}$ external forces. The porous medium is modeled using the Lorentz gas (or `Swiss-cheese model')~\cite{Halperin:1985}, composed of $M$ randomly distributed, overlapping discs of radius $a$ in a square domain of width $H$. We further add empty patches to the channel inlet and outlet so that the resulting domain is a rectangular channel of length $L=1.4H$ and width $H$ (Fig.\ref{fig:set-up}(a)). Fluid can move through the free space between the discs, denoted by $\Omega$. We impose no-slip boundary conditions (BC) $\vec{u}=\vec{0}$ at $\partial\Omega$ and periodic BCs along~$\vec{e}_2$. Further, we apply a constant force (density) at the inlet of the channel $\vec{f}=f\delta_{\rm in}(\vec{r})\vec{e}_1$, where $f >0$ and $\delta_{\rm in}$ is the Kronecker-delta identifying the inlet of the channel. This leads to a macroscopic pressure drop along the channel ($\vec{e}_1$), $\Delta p = p_{\rm inlet}- p_{\rm outlet}$, hence mimicking pressure-driven flow. By using $a$ as relevant length scale, $fa/\mu$ as characteristic velocity, and $f$ as pressure scale, we identify one non-dimensional parameter: $H/a$. We solve Eqs.~\eqref{eq:stokes} numerically using a finite-element scheme based on Taylor-Hood elements and properly adapt the mesh close to tight spaces and sharp corners~\cite{supp}.

The porosity of the Lorentz gas is characterized by the packing fraction of the discs: $\phi=1-\exp(-M\pi a^2/H^2)$~\cite{Torquato:book}. In percolation theory, an important quantity represents the probability that two points at a given distance $r$ belong to the same cluster, which is encoded in the pair-connectedness function $p_2(r)\sim \exp(-r/\xi)$~\cite{Grimmett:1989} with correlation length $\xi$ as relevant length scale. As the percolation threshold (i.e., the density of obstacles $\phi_c$ at which a sample-spanning cluster first appears) is approached, it is predicted to obey $\xi\sim (\phi_c-\phi)^{-\nu}$ with critical packing fraction $\phi_c\approx0.67637$~\cite{Quintanilla:2000, Quintanilla:2007} and universal exponent $\nu=4/3$ in 2D~\cite{Grimmett:1989,Torquato:book}. In our work, we consider systems of width $H=[12.5,150]a$ and use packing fractions of $0.05\phi_c$ to $0.98\phi_c$. Quantities, such as the flow rate and velocity distributions, are obtained by averaging over at least 30 different statistically independent geometries~\cite{supp}. 

\begin{figure}[tp]
\includegraphics[width=\linewidth]{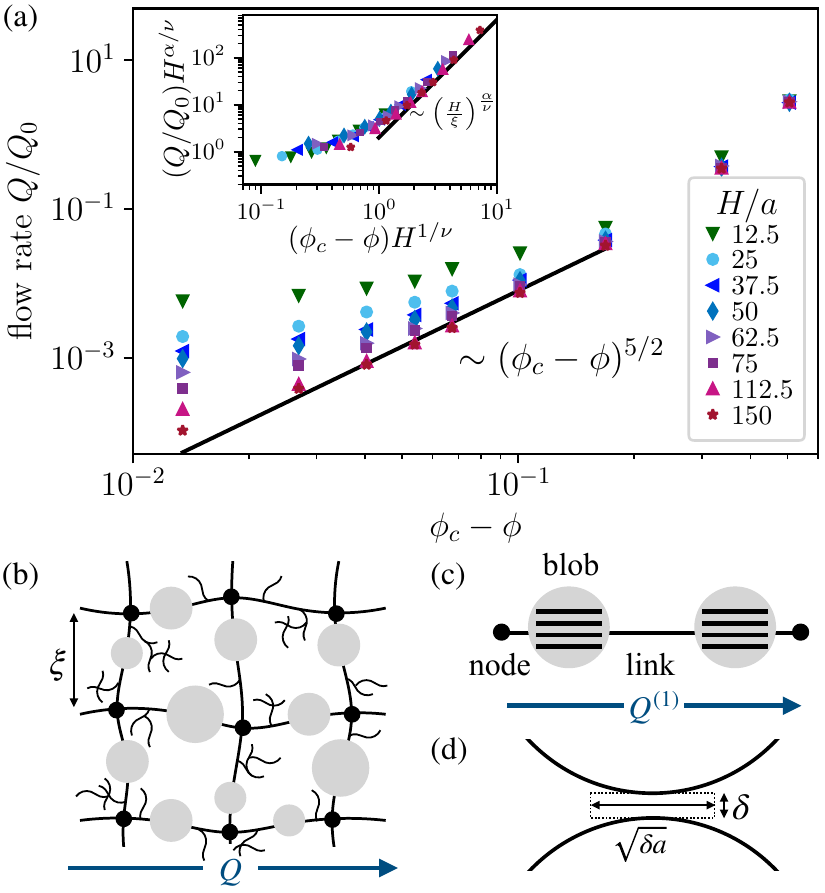}
\caption{(a) Flow rate $Q$ as a function of the packing fraction $\phi_c-\phi$ for different system sizes $H/a$. The solid line inidicates the scaling prediction. The flow rate is rescaled by $Q_0=f/(\mu H)$. ({\it Inset})~Finite-size scaling of~$Q$ following Eq.~\eqref{eq:FSS}. Here, $\xi\sim(\phi_c-\phi)^{-\nu}$ denotes the correlation length. (b-c) Sketch of a flow network composed of nodes connected through links and blobs. The tangling ends correspond to dead-end-channels of the porous medium and constitute the dominant part of the void space. The flow rate between two nodes is denoted by $Q^{(1)}$. (d) Sketch of a characteristic channel of the porous medium having a width $\delta$ and length $\sqrt{\delta a}$. \label{fig:flow_rate}}
\end{figure}

\paragraph{Power-law scaling of the flow-rate.--} Figure~\ref{fig:set-up} displays the spatially varying pressure field $p(\vec{r})$ in a highly heterogeneous, porous environment. We observe large regions of constant pressure, whose magnitude decreases from the channel inlet to its outlet (Fig.\ref{fig:set-up}(a)). These regions result from very few open pores that enable fluid flow from one area of the pore space to another and can extend across the entire channel width. Consequently, this pressure field then leads to the formation of very few streams of high velocities through the open channels of the porous medium (Figs.\ref{fig:set-up}(b-c)), forming the flow backbone (Fig.\ref{fig:set-up}(d)), and large stagnant zones of vanishing flow in the areas of constant pressure.

This highly-heterogeneous flow field $\vec{u}=(u_1,u_2)$ can be quantified in terms of the macroscopic flow rate via $Q = \int_0^H u_1(x,y) \diff y$. In practice, we also average over the length of the porous domain $x\in [0,H]$. Strikingly, we observe a power-law scaling of the flow rate as the percolation threshold $\phi_c$ is approached, $Q\sim (\phi_c-\phi)^\alpha$. Fitting the data for $H/a=150$ and $\phi_c-\phi\lesssim 0.02$, yields an exponent of $\alpha \simeq 2.44\approx 5/2$ (Fig.\ref{fig:flow_rate}(a)). 

Our findings agree with the scaling exponent previously predicted by Halperin et al.~\cite{Halperin:1985}. In their approach, the void space of the porous medium is mapped to a discrete random network (the `nodes-links-blobs' model~\cite{Coniglio:1982}) composed of {\it nodes} which are separated at a distance $\sim \xi$ and connected through sequences of {\it links} (composed of connected channels) and {\it blobs} (a group of connected channels in parallel), see Figs.\ref{fig:flow_rate}(b-c). As dead-end channels do not contribute to transporting fluid across the medium, we do not consider them here. To obtain the flow rate - pressure drop  ($Q^{(1)} -\Delta p$) relation between two nodes, we ignore the resistance due to the blobs and compute the flow through a single link. Therefore, we first write the pressure drop as sum of pressure drops $\Delta p_i$ of $N$ individual channels: $\Delta p = \sum_{i=1}^N \Delta p_i$. Taking into account the geometry of a narrow channel of width $\delta_i$ composed of two opposed discs of radius $a$, one identifies its length as $\sqrt{a\delta_i}$ (Fig.\ref{fig:flow_rate}(d)), leading to
\begin{align}
\Delta p_i &= 12 Q^{(1)} \mu \delta_i^{-5/2}a^{1/2}. \label{eq:qi}
\end{align}
To evaluate the resulting relation,
\begin{align}
\Delta p & \sim Q^{(1)} \sum_{i=1}^{N} \delta_i^{-5/2}, \label{eq:Q1_sum}
\end{align}
a few observations are of importance. First, the bonds~$\delta_i$ of a Lorentz gas obey a distribution $\varphi(\delta)$, which is finite as $\delta \to 0^+$ \cite{Torquato:book, Spanner:2016}. Next, the number of channels~$N$ increases with increasing $\phi$ and scales as $N\sim (\phi_c-\phi)^{-1}$ for two points separated at a distance $\sim \xi$~\cite{Coniglio:1982}. These aspects allow us to replace Eq.~\eqref{eq:Q1_sum} by an integral
\begin{align}
\Delta p &\sim Q^{(1)}N\int_{\delta_0}^{\infty}\varphi(\delta)\delta^{-5/2}\diff\delta,
\end{align}
where $\delta_0$ denotes the minimum bond width of the string. As the width of the channels $\delta$ becomes smaller for increasing $\phi$, we can assume $\delta_0\approx 1/(\varphi(0)N)$ (for large $N$) and consequently find that the dominating contribution stems from the weakest bond~\cite{Halperin:1985}, leading to $Q^{(1)}\sim N^{-5/2} \sim (\phi_c-\phi)^{5/2}$. Finally, the macroscopic flow rate $Q$ is related to $Q^{(1)}$ in 2D by considering a network of $L/\xi$ links connected in series and $L/\xi$ of the series of links in parallel: 
\begin{align}
Q\sim (\phi_c-\phi)^{5/2}.
\end{align}
Most importantly, our excellent agreement with numerical simulations suggests that the scaling behavior of the flow rate $Q$ is fully determined by a flow network, which requires only knowledge of flow through a single narrow channel and geometric features of the Lorentz model. Thus it is quite remarkable that despite the complexity of our system (see Fig.\ref{fig:set-up}), our numerical measurements of the flow rate are indeed explained by the scaling theory~\cite{Halperin:1985}, where, for instance, the contribution of blobs is ignored, and,
 analytical predictions for channel flow in a simple geometry suffices to describe the macroscopic flow rate in a dense porous environment.  

\begin{figure}[tp]
    \centering
    \includegraphics[width=1.0\linewidth]{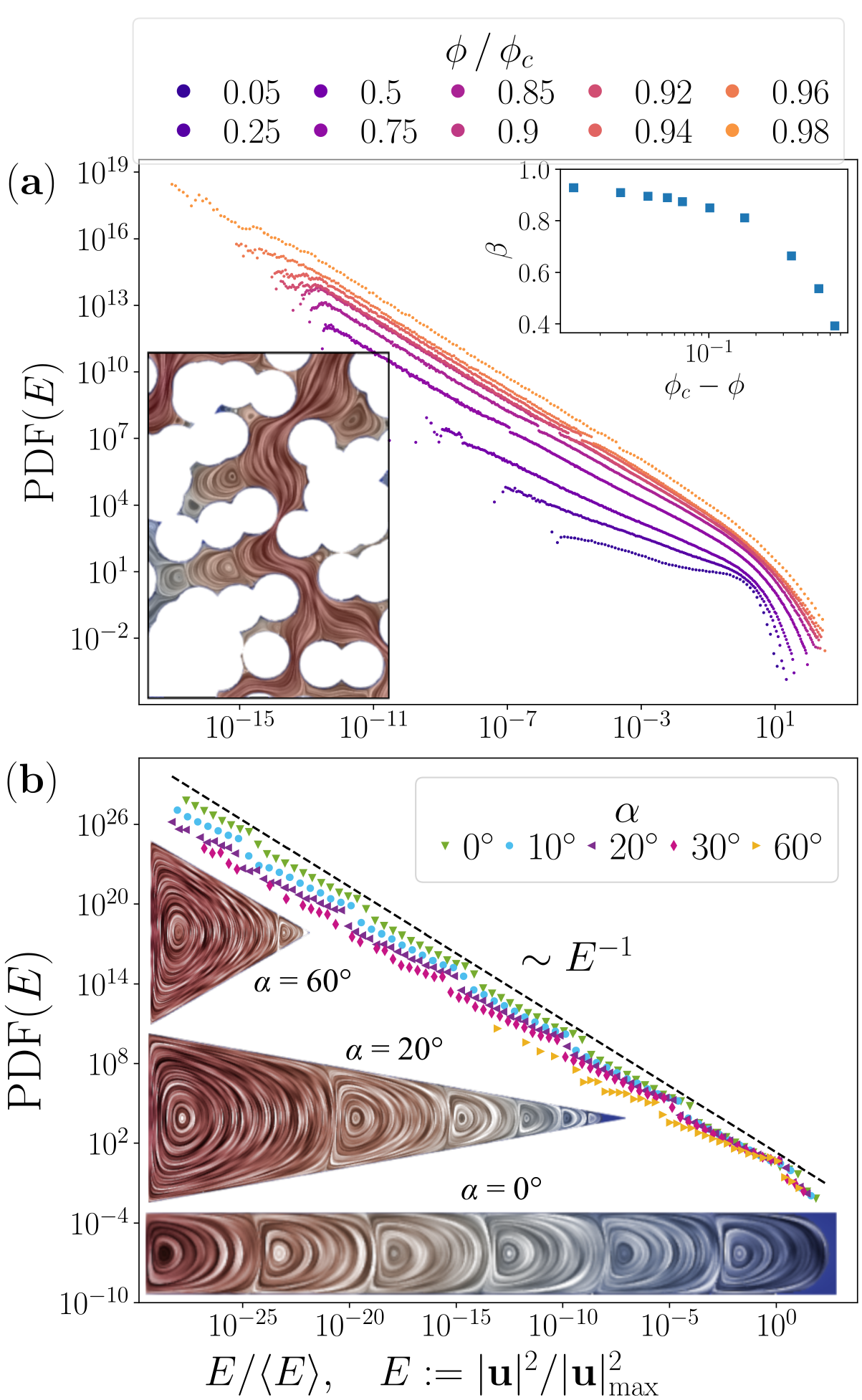}
    \caption{(a)~Probability density function (PDF) of the rescaled kinetic energy, $E=|\vec{u}|^2/|\vec{u}|_{\rm max}^2$, where $|\vec{u}|_{\rm max}$ denotes the maximal velocity. ({\it Top right})~Exponent $\beta$ of the power-law behavior for small kinetic energies. ({\it Bottom left})~Example of a stagnant zone with eddies in the dead-end-pores induced by the tangent velocity of the backbone flow (dark-red).  
    (b)~${\rm PDF}(E)$ for corner flows in different geometries. ({\it Inset})~Collection of viscous eddies emerging in different channel geometries. Here, $\alpha$ denotes the angle between the two long sides of the channels, where $\alpha=0^{\circ}$ represents the limiting case of parallel plates. 
    The height of the channel is kept constant for different $\alpha$. The channels are re-scaled for visualization purposes. The magnitudes of the velocities are shown in a logarithmic scale, with red corresponding to $\sim 1$ and dark blue to $10^{-15}$.} 
    \label{fig:pdf}
\end{figure}

{\it Finite-size scaling.--} We finally stress that this critical behavior in principle holds for infinitely large systems ($H\to\infty$) and is expected to be recovered for finite system sizes that are larger than the correlation length of the void space, $H\gtrsim\xi$.  Our results show that deviations appear for smaller system sizes, $H/a\lesssim 100$, where the curves of the flow rate bend and the power-law behavior becomes rounded. To interpret these results we perform a finite-size scaling ansatz. In particular, we note that the scaling prediction for the flow rate can be expressed by $Q\sim \xi^{-\alpha/\nu}$. For small systems, $H\ll\xi$, $H$ assumes the role of $\xi$, modifying the prediction to $Q\sim H^{-\alpha/\nu}$ for $\phi\to\phi_c$. Thus, we use the following ansatz~\cite{Torquato:book, Binder:2010} 
\begin{align}
Q/Q_0 &= H^{-\alpha/\nu}f(H^{1/\nu}(\phi_c-\phi)), \label{eq:FSS}
\end{align}
with scaling function $f(x) = {\rm cte.}$ for $H\ll \xi$ and $f(x) \sim x^\alpha$ for $H\gg \xi$. This ansatz indeed nicely collapses our data covering sizes of $12.5\leq H/a\leq 150$  at high packing fractions (Fig.\ref{fig:flow_rate}(a)(inset)). Our results thus suggest that $\xi$ is the relevant length scale of our system, and the expected scaling with exponent $\approx 5/2$ is only smeared out by finite-size effects.    

\paragraph{Kinetic energy.--} We further analyze the distribution of the kinetic energy, denoted as $E = |\vec{u}|^2/|\vec{u}|_{\rm max}^2$ (Fig.\ref{fig:pdf}(a)). As the packing fraction increases, we observe a noteworthy power-law scaling behavior, ${\rm PDF}(E)\sim E^{-\beta}$, spanning several decades in the range of $\sim 10^{-9}-10^{-1}E$, with the exponent $\beta$ approaching $\approx0.9$. This power-law scaling emerges at small kinetic energies, which can be primarily found in the dead-end-pores of the medium (i.e. the tangling ends of the random resistor network). Within these regions, the flow consists of collections of viscous eddies of varying sizes (Fig.\ref{fig:pdf}(a)(bottom left)). To rationalize whether the eddies contribute to the power-law scaling of ${\rm PDF}(E)$, we investigate Stokes flow across different channel geometries. These reveal collections of eddies (Fig.\ref{fig:pdf}(b)) with a significant reduction in velocities (and consequently energies) spanning orders of magnitude, consistent with predictions for viscous corner flows in terms of asymptotic, self-similar solutions~\cite{Moffatt:1964}. The resulting PDFs exhibit a power-law scaling with a geometry-dependent exponent of $\beta \sim 1$ for straight channels ($\alpha=0^\circ$) that decreases for triangular shapes ($\beta \sim 0.8$ for $\alpha = 60^\circ$). Notably, these exponents closely resemble those observed in the porous medium, implying that the ${\rm PDF}(E)$ at low $E$ is predominantly influenced by a collection of such viscous eddies within dead-end channels. Thus, we anticipate that the power-law behavior emerges due to the self-similar nature of the individual viscous eddies and the extended power-law represents a collection of these. 

\paragraph{Conclusions.--} Our findings on Stokes flow through porous media demonstrate a power-law scaling of the macroscopic flow rate with an exponent $\approx 5/2$ at the percolation threshold, thus confirming, for the first time, analytical predictions by Halperin et al.~\cite{Halperin:1985}. We observe that in very dense environments, the underlying flow field becomes highly heterogeneous, characterized by a few fast streams transporting fluid across the porous medium. This feature allows mapping the behavior of the macroscopic flow to a discrete flow network, whose structure relies on geometric features of the Lorentz gas~\cite{Halperin:1985}. Moreover, our study has provided a firm foundation for understanding Stokes flow through dense porous media via a finite-size scaling analysis.

Looking ahead, our findings lay the foundation to study the physics of complex fluids in disordered media and unraveling power-law behaviors for e.g., non-Newtonian~\cite{Datta:2022, Kumar:2022} or active fluids~\cite{Wioland:2013, Wioland:2016, Secchi:2016,Doostmohammadi:2018, Keogh:2024,Jorge:2024}. In particular, scaling relations for their critical behavior may be derived for stationary, non-Newtonian flows using analytical predictions for their flow rate through narrow channels as input~\cite{Boyko:2021,Boyko:2022}, in a regime where memory effects can be ignored. We anticipate that testing these predictions through numerical and experimental means will enhance our understanding of non-Newtonian flows through disordered media, potentially utilizing substances like Xanthan gum solution~\cite{Pipe:2008}, for experimental validation.

While our focus has been on Stokes flow through the Lorentz gas, we believe our results can be extended to other porous media models, especially those close to percolation. These models, which could feature different channel shapes~\cite{Scholz:2012} and pore size distributions, may similarly be mapped onto the `nodes-links-blob' network.

We further note that transport of finite-sized particles through porous media can significantly change the properties of the velocity field~\cite{Residori:2023} and so the flow rate can be used as measure for the dynamic properties of natural porous media, such as sediments.

Finally, several (semi-)empirical expressions for the permeability, introduced via Darcy's law, have been derived to quantify and predict flow through porous media. These often depend on the conductivity of the fluid phase and microscopic parameters~\cite{Katz:1986, Scholz:2012} (e.g., the Katz-Thompson relation). However, previous work pointed out that the flow rate is not related to the conductivity due to the different nature of the Laplace and Stokes equations~\cite{Torquato:book}. Our study now resolves this aspect and offers valuable guidance for future experiments, particularly through a comprehensive finite-size analysis.

\paragraph{Acknowledgements.--} C.K. acknowledges helpful discussions with Evgeniy Boyko, Thomas Franosch, Howard~A. Stone, Alexander Wietek, and Akhil Varma. The authors gratefully acknowledge computing time granted by the Center for Information Services and High-Performance Computing [Zentrum für Informationsdienste und Hochleistungsrechnen (ZIH)] at the TU Dresden.

\bibliography{literature}

\end{document}



\title{Supplemental Material: Flow Through Porous Media at the Percolation Transition}

\author{Mirko Residori}
\affiliation{Institute of Scientific Computing, Technische Universit\"at Dresden, 01062 Dresden, Germany}
\affiliation{Max Planck Institute for the Physics of Complex Systems, N\"othnitzer Stra{\ss}e 38,
01187 Dresden, Germany}
\author{Suvendu Mandal}
\affiliation{Technische Universit\"at Darmstadt, Karolinenplatz 5, 64289 Darmstadt, Germany}
\author{Axel Voigt}
\affiliation{Institute of Scientific Computing, Technische Universit\"at Dresden, 01062 Dresden, Germany}
\affiliation{Center for Systems Biology Dresden, Pfotenhauerstraße 108, 01307 Dresden, Germany}
\affiliation{Cluster of Excellence, Physics of Life, TU Dresden, Arnoldstraße 18, 01307 Dresden, Germany}
\author{Christina Kurzthaler}
\email{ckurzthaler@pks.mpg.de}
\affiliation{Max Planck Institute for the Physics of Complex Systems, N\"othnitzer Stra{\ss}e 38,
01187 Dresden, Germany}
\affiliation{Center for Systems Biology Dresden, Pfotenhauerstraße 108, 01307 Dresden, Germany}
\affiliation{Cluster of Excellence, Physics of Life, TU Dresden, Arnoldstraße 18, 01307 Dresden, Germany}

\maketitle

In the supplementary material we present a description of the numerical method, details of the data analysis, and additional results, such as distributions of velocities and the tortuosity. 

\section{Detailed description of the numerical method}
Let $\mathtt{Rect}$ be a rectangular domain and $\cup_{i=1}^{N} D_i$ a collection of $N$ possibly overlapping discs of radius $a$. We consider the domain $\Omega = \mathtt{Rect}\setminus \cup_{i=1}^{N} D_i$. 
Let $\mathtt{Rect}_{L,R,T,B}$ be the left, right, top, and bottom edges of $\Omega$, respectively. We set the inlet of the channel at $\mathtt{Rect}_L$. Further, we define the space $\vec{V} := \{ \vec{v}\in H^1(\Omega)^2 : \vec{v} = \vec{0} \text{ on } \partial\left\{\cup_{i=1}^N D_i\right\}, \, \nabla\vec{u}\cdot\vec{n} = \vec{0} \text{ on } \mathtt{Rect}_{L,R}, \, \vec{u} \text{ periodic on } \mathtt{Rect}_{T,B}\}$ and consider weak solutions $(\hat{\vec{u}},\hat{p})\in\vec{V}\times L^2(\Omega)$ of 
\begin{align}
\mu \nabla^2 \vec{u} - \nabla p + \vec{f}=\vec{0} \quad \text{and} \quad
\nabla\cdot\vec{u} = 0, \label{eq:stokes}
\end{align}
such that 
\begin{subequations}
\begin{align}
\into \nabla \vec{v} : \nabla \hat{\vec{u}} - \hat{p}\nabla\cdot\vec{v}\,\mathrm{d}\vec{x} &=  \into \vec{f}\cdot \vec{v}, \\
\into q\nabla\cdot\hat{\vec{u}}\,\mathrm{d}\vec{x}&=0,
\end{align}
\end{subequations}
for every $\vec{v} \in H^1_{\vec{0}}(\Omega)^2$ and  $q\in L^2(\Omega)$. We partition the domain $\Omega$ by a conforming triangulation $\mathcal{T}_h$. Then, the continuous spaces $\vec{V}\times L^2$ are approximated by the Taylor-Hood space $\mathbb{T}_h$  defined as 
\begin{align*}
\mathbb{T}_h &= (\mathbb{V}_{2,\mathrm{BC}}\times \mathbb{V}_{2,\mathrm{BC}}) \times \mathbb{V}_1, \\
\mathbb{V}_{m,\mathrm{BC}} &= \{v\in \mathbb{V}_m: v = \vec{0} \text{ on } \partial\{\cup_{i=1}^N D_i\}, v \text{ periodic on }\mathtt{Rect}_{T,B}\},\\
\mathbb{V}_m &= \{p\in C^{0}(\Omega) : p|_{T} \in \mathbb{P}_m(T), \forall T\in\mathcal{T}_h\},
\end{align*}
where $\mathbb{P}_m$ is the space of polynomials of order at most $m$. The velocity $\vec{\hat{u}}$ and pressure $\hat{p}$ are approximated by functions from $\mathbb{T}_h$, i.e. $m=2$ piece-wise quadratic for $\vec{\hat{u}}$ and $m=1$ piece-wise linear for $\hat{p}$. Particular attention has to be paid to the choice of the triangulation $\mathcal{T}_h$: as discs are randomly generated and are allowed to overlap, narrow regions between the discs are likely to appear within the domain $\Omega$. Each of these regions should be properly resolved. To this end we adapt the mesh in close proximity to tight spaces. Mesh generation is performed with \texttt{GMSH}~\cite{gmsh}. The numerical solution is realized using the finite element toolbox \texttt{AMDiS} \cite{vey2007amdis,witkowski2015software} which is based on the \texttt{DUNE} framework~\cite{dune, alugrid}. The resulting linear system is solved by the \texttt{UMFPACK} solver provided  by the \texttt{SuiteSparse} library~\cite{umfpack}.

\section{Detailed data processing} 
In Fig.~3 of the main text and in the subsequent section we report several probability density functions (PDF) related to the velocity field $\vec{u}$. The bins of the PDFs for Fig.~3 in the main text are log-spaced and the number of bins is chosen as the square root of the number of data. Differently, the bins in the subsequen section are constructed by following the Freedman-Diaconis rule, i.e. by choosing the bin width $\Delta x$ as
\[
\Delta x = 2\frac{\mathrm{IQR}(\mathrm{data})}{\sqrt[3]{N}},
\]
where $N$ is the size of $\mathrm{data}$ and $\mathrm{IQR}$ is the interquartile range, i.e. the difference between the 75\textit{th} and 25\textit{th} percentile of $\mathrm{data}$. 
The data are considered above a threshold value: $\vec{u}$ where $|\vec{u}| / \langle|\vec{u}|\rangle \geq 10^{-4}$. Then, data are binned with a corresponding weight resulting in a histogram. The corresponding weight is related to the unstructured grid on which the velocity field $\vec{u}$ is defined. For every element of the grid we compute an average value of the vector field associated to that element. The corresponding weight is the area of the element. We ignore bins containing a negligible number of points in the histogram. Finally, we normalize the histogram to be a PDF. Each PDF is obtained by collecting data over at least 35 independent realizations of the geometry for each packing fraction $\phi$ in a single array. Statistics are then computed over the whole data set. 

Here, the average fluid velocity is computed as 
\[
\langle|\vec{u}|\rangle = \frac{\int_{\Omega_o} |\vec{u}|\mathrm{d}\vec{x}}{|\Omega_o|},
\]
where $\Omega_o$ is the connected pore-space, i.e.,  the void space between the obstacles. We further extract velocities that are in magnitude larger than the average fluid velocity, representing the portion of the porous-media in which there is an appreciable flow. This subset is also known as the backbone of the porous-media, see Fig.(1)(d) in the main text and Fig.\ref{fig:highphi}(c) for examples.

\section{Additional results\label{sec:additional_results}}
\begin{figure*}[htp]
\begin{tikzpicture}
    \node[inner sep=0pt] at (-10.4,0) {\includegraphics[width=0.46\linewidth, trim={245 205 245 205}, clip]{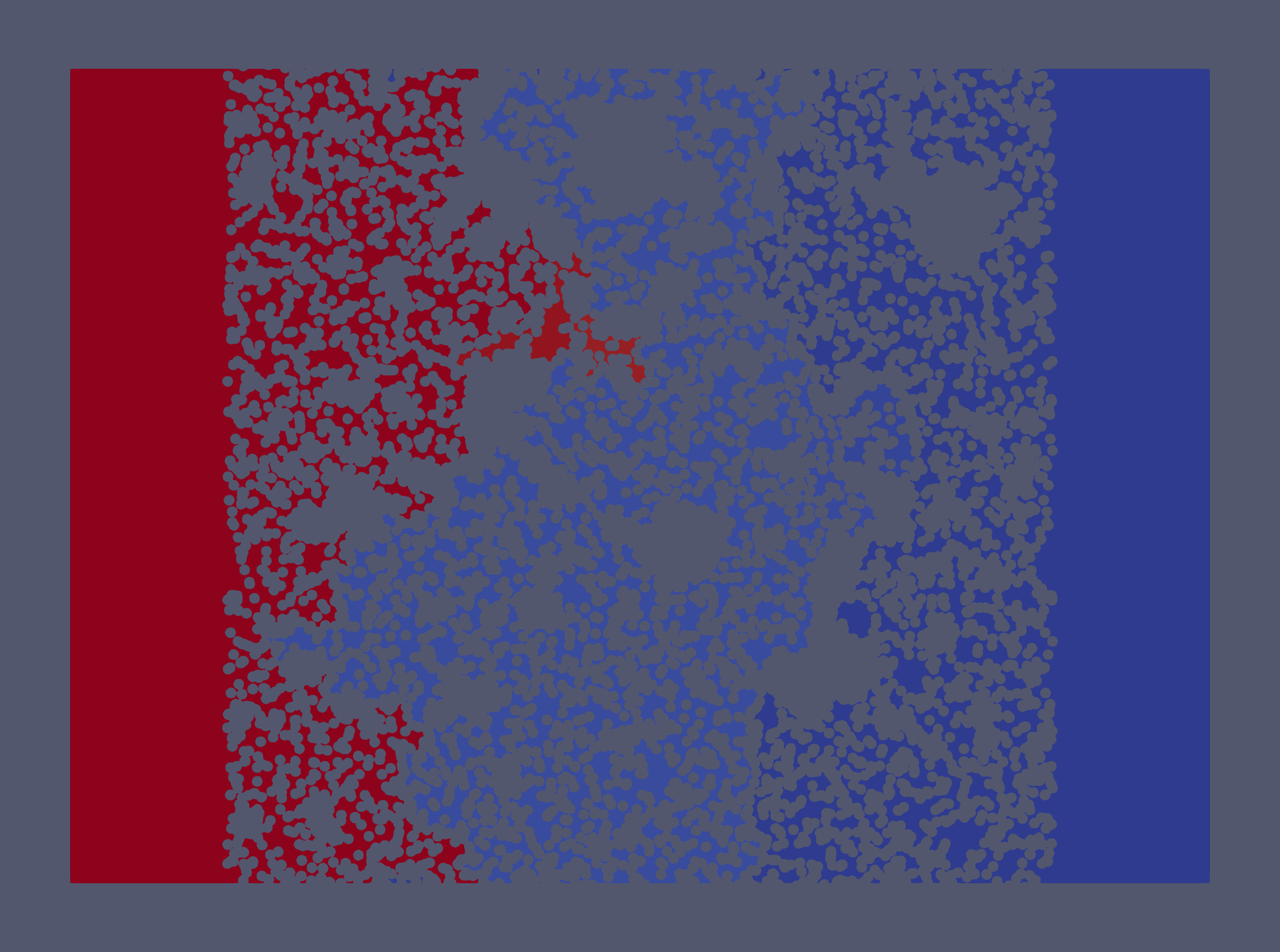}};
    \draw[white, very thick, <->] (-13.4,-2.95) -- (-13.4,2.95) node [midway, above, sloped] {$H$};
    \draw[black, very thick, <->] (-14.5,-3.1) -- (-6.3,-3.1) node [midway, below, sloped] {$L$};
    \node[white, rotate=90] at (-14.3,1.5) {inlet};
    \node[white, rotate=90] at (-6.5,1.5) {outlet};
    \draw[white, very thick, ->] (-14.5, -1.0) -- (-13.7, -1.0) node [midway, above] {$\mathbf{f}$};
    \draw[white, very thick, ->] (-14.3, -2.8) -- (-13.8, -2.8) node [pos=0.95, above] {$\vec{e}_1$};
    \draw[white, very thick, ->] (-14.3, -2.8) -- (-14.3, -2.3) node [pos=0.95, above] {$\vec{e}_2$};
    \node at (-7.0, -0.9) {
    \pgfplotscolorbardrawstandalone[ 
    colorbar,
    point meta min=0,
    point meta max=1.0,
    colorbar style={
           tickwidth=2pt,
            title=$p(\vec{r})/f$,
            text=white,
            height=3cm,
            width=0.3cm,
            ytick={0.0,.2,0.4,0.6,0.8,1.0}         
            }
    ]
    };
    \node[inner sep=0pt] at (-3.2,-0.15) {\includegraphics[width=0.34\linewidth, trim={375 120 375 120}, clip]{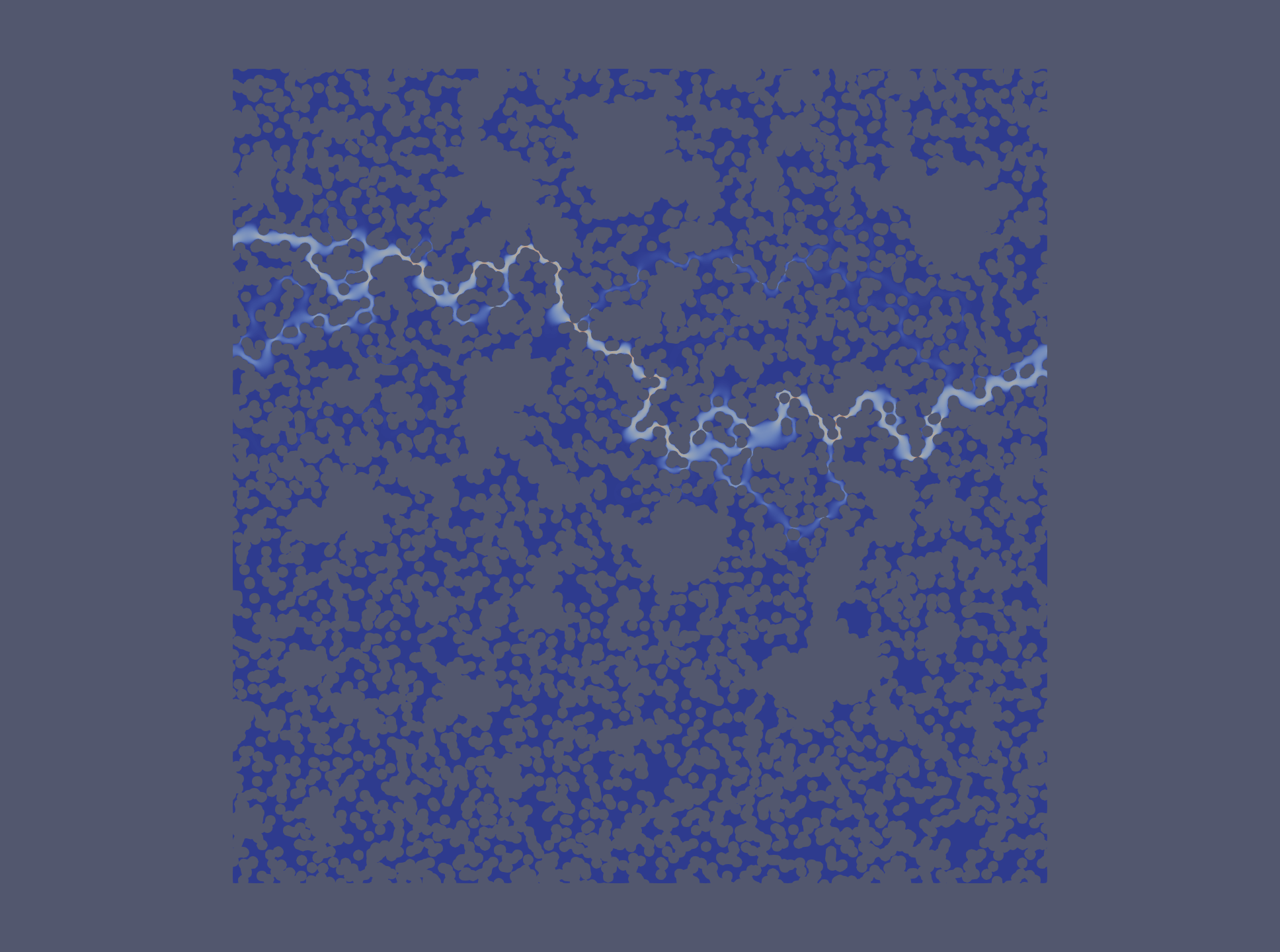}};
    \node at (-0.6, -1.6) {
    \pgfplotscolorbardrawstandalone[ 
    colorbar,
    point meta min=0,
    point meta max=8.0,
    colorbar style={
           tickwidth=2pt,
            text=white,
            height=3cm,
            width=0.3cm,
            ytick={0.0,2.0,4.0,6.0,7.9}         
            }
    ]
    };
    \node[white, rotate=90] at (-1.3,-1.7) {$\log(1+|\mathbf{u}|/\langle |\mathbf{u}|\rangle)$};  
    \node[inner sep=0pt] at (1.6,-0.15) {\includegraphics[width=0.191\linewidth, trim={650 150 650 150}, clip]{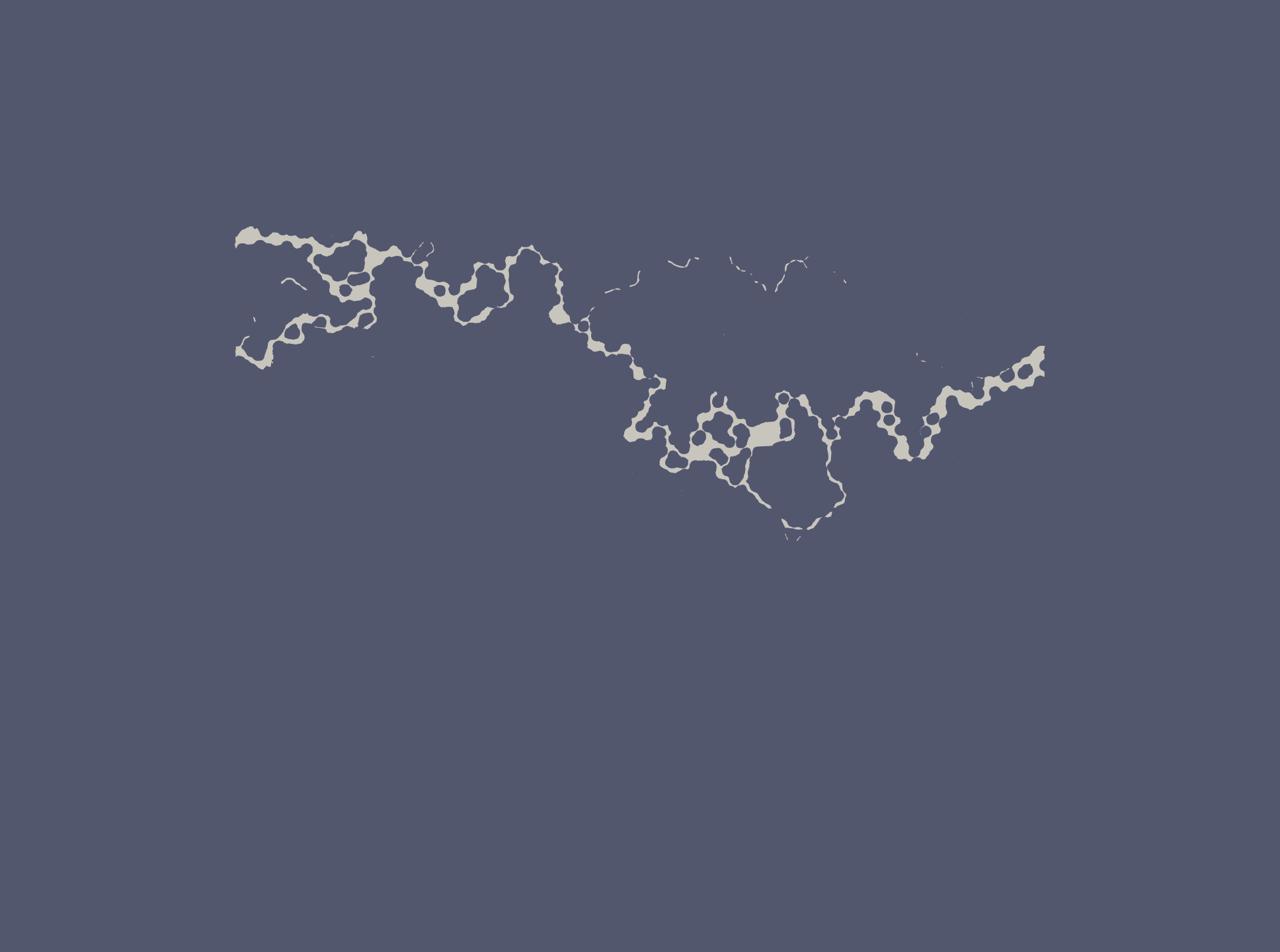}};
    \node[white, very thick] at (-14.2,2.7) {\textbf{(a)}};
    \node[white, very thick] at (-5.9,2.7) {\textbf{(b)}};
    \node[white, very thick] at (0.2,2.7) {\textbf{(c)}};
\end{tikzpicture}
\caption{(a) Pressure field $p(\vec{r})/f$. Here, the width and length are $H=150a$ and $L = 1.4H$, respectively, where $a$ denotes the disc radius. The packing fraction of overlapping discs is $\phi\approx 0.98\phi_c$, where $\phi_c$ is the critical packing fraction. We consider periodic BC along the vertical direction. Disconnected void space is shown in grey. (b)~Velocity field. Rescaled magnitude of the velocity $|\vec{u}| / \langle|\vec{u}|\rangle$ is shown in a logarithmic scale, where $\langle|\vec{u}|\rangle$ is the velocity magnitude averaged over the connected part of the sample. (c) Part of the backbone of the velocity field, where regions of velocities $\ge \langle|\vec{u}|\rangle$ are indicated in light grey. 
\label{fig:highphi}}
\end{figure*}

We present the pressure and flow fields for a packing fraction of $\phi = 0.98\phi_c$ in Fig.\ref{fig:highphi}. At this high packing fraction, only few streams transporting fluid from the channel inlet to its outlet are observed.

\begin{figure}[htp]
\centering
\includegraphics[width=1.0\textwidth]{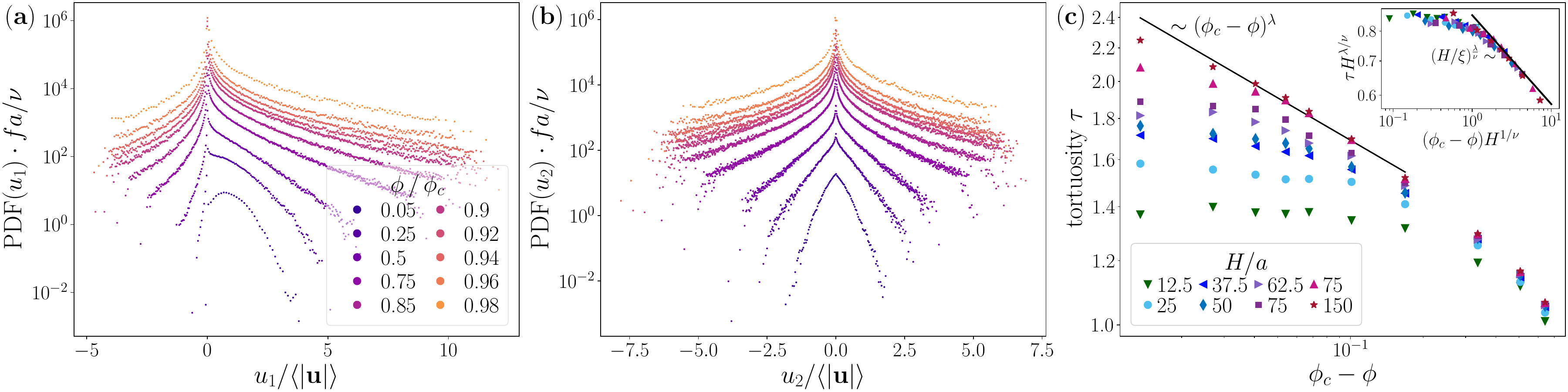}
\caption{Probability density function (PDF) of velocities (a) along ($u_1$) and (b) perpendicular ($u_2$) to the direction of the applied forcing for different packing fractions $\phi$. The $x-$axis is rescaled by the average velocity $\langle |\vec{u}|\rangle$. Results are shown for $H/a=150$. (c) Tortuosity $\tau$ as a function of the rescaled packing fraction $\phi_c-\phi$ for different system sizes $H/a$. Here, $\phi_c$ denotes the critical packing fraction. ({\it Inset}) Finite-size scaling of the tortuosity, following Eq.(6) of the main text, with $\lambda \approx -0.17$. \label{fig:additional}}
\end{figure}

To further characterize the flow, we compute the distributions for flow velocities along, $u_1$, and perpendicular, $u_2$, to the applied forcing. At low packing fraction, $0.05\phi_c$, the distribution for $u_1$ is anisotropic exhibiting a peak at  $u_1\sim\langle|\vec{u}|\rangle$ and no back flow, as expected in dilute environments (Fig.\ref{fig:additional}(a)). As the packing fraction increases back flows emerge, contributing to the probability of negative velocities, and the peak shifts to $u_1/\langle|\vec{u}|\rangle\to0$, as indicator for the dominating contribution of stagnant zones. A distinct feature at high packing fractions are long tails in the probability density, which reflect large variations of the velocities appearing in the few fast streams. The distributions of $u_2$ are symmetric and peaked at $u_2/\langle|\vec{u}|\rangle=0$ (Fig.\ref{fig:additional}(b)). For small $\phi$ the tails of the distributions appear exponential, however, the shape of the peak becomes sharper as $\phi$ increases and long tails emerge, which is in qualitative agreement with Ref.~\cite{Matyka:2016}.

Finally, we have computed the tortuosity, via $\tau = \langle |\vec{u}| \rangle /\langle u_1\rangle$~\cite{Duda:2011}, which measures the average elongation of fluid flow. It shows a scaling with exponent $\lambda \approx -0.17$ for the largest system size, in agreement with previous work~\cite{Duda:2011}. We further find a full data collapse  using a finite-size scaling ansatz~\cite{ghanbarian:2022} as in Eq.~(6) of the main text for all packing fractions and system sizes, see Fig.~\ref{fig:additional}.

\bibliography{literature}